# Human capital in the sustainable economic development of the energy sector

**Evgeny Kuzmin [1], Maksim Vlasov [2, 3], Wadim Strielkowski [4,*], Marina Faminskaya [6] and Konstantin Kharchenko [7]**

[1] Department of Regional Industrial Policy and Economic Security, Institute of Economics of the Ural Branch of the Russian Academy of Sciences, Moskovskaya str. 29, 620014 Ekaterinburg, Russia; kuzmin.ea@uiec.ru
[2] Center of Economic Theory, Institute of Economics of the Ural Branch of the Russian Academy of Sciences, Moskovskaya str. 29, 620014 Ekaterinburg, Russia; vlasov.mv@uiec.ru; mv.lassov@mail.ru
[3] Department of Regional Economics, Innovative Entrepreneurship and Security, Ural Federal University, Mira str. 19, 620002 Ekaterinburg, Russia; vlasov.mv@uiec.ru
[4] Department of Trade and Finance, Faculty of Economics and Management, Czech University of Life Sciences Prague, Kamýcká 129, Prague 6, 165 00 Prague, Czech Republic
[5] Department Agricultural and Resource Economics, University of California, Berkeley, 303 Giannini Hall, CA 94720, United States
[6] Faculty of Information Technology, Russian State Social University, Wilhelm Pik str. 4/1, 129226 Moscow, Russia; faminskaya@ymservices.ru
[7] Departments of State and Municipal Administration, Financial University under the Government of the Russian Federation, Leningradsky Prospekt 49, 125167 Moscow, Russia, Russia; geszak@mail.ru; k.kharchenko.k@inbox.ru

* Correspondence: strielkowski@berkeley.edu

**Abstract:** This study examines the role of human capital investment in driving sustainable socio-economic growth within the energy industry. The fuel and energy sector undeniably forms the backbone of contemporary economies, supplying vital resources that underpin industrial activities, transportation, and broader societal operations. In the context of the global shift toward sustainability, it is crucial to focus not just on technological innovation but also on cultivating human capital within this sector. This is particularly relevant considering the recent shift towards green and renewable energy solutions.

In this study, we utilize bibliometric analysis, drawing from a dataset of 1933 documents (represented by research papers, conference proceedings, and book chapters) indexed in the Web of Science (WoS) database. We conduct a network cluster analysis of the textual and bibliometric data using VOSViewer software. The findings stemming from our analysis indicate that investments in human capital are perceived as important in achieving long-term sustainable economic growth in the energy companies both in Russia and worldwide. In addition, it appears that the role of human capital in the energy sector is gaining more popularity both among Russian and international researchers and academics.

**Keywords:** energy sector; sustainable development; human capital; renewable energy sources; Russia

## 1. Introduction

Human capital embodies the array of competencies, insights, talents, and characteristics that people in a society or labor force collectively hold. It includes formal academic and vocational education as well as the tacit understanding gained from job experiences [1, 2]. This non-physical resource is essential in propelling the socio-economic progress of organizations within the energy sector. Moreover, capitalizing on human capital substantially influences the evolution and prosperity of areas predominantly driven by the energy and fuel industry [3, 4]. The reliance of energy sector entities on adept individuals is critical for fostering innovation, adjusting to new technologies, and enhancing overall efficiency and creativity [5]. Investments in the intellectual growth, expertise, and health of employees through various professional development endeavors enrich the human capital of these corporations [6]. A proficient labor force exerts a positive impact not only on a company's output but also on the socio-economic fabric of a region. Competent workers draw further investment from diverse sectors seeking skilled labor, which results in job creation, higher earnings, improved quality of life, poverty reduction, greater social advancement, and, ultimately, enduring economic progress [7]. Additionally, investment in human capital is instrumental in nurturing innovation and entrepreneurial ventures, with skilled individuals more inclined to initiate business ventures that promote economic variety and new commercial possibilities [8].

Hence, the energy sector's role in the socio-economic development of locales where energy firms are situated is substantial. This broad sector, which includes the procurement, refinement, production, and distribution of fossil fuels such as oil, gas, and coal, offers both prospects and challenges for the regions it serves [9, 10]. One of the foremost outcomes of human capital investment in this complex is the generation of job opportunities, as these industries demand a skilled workforce to function optimally [11]. Such investment is advantageous for individuals through the provision of secure employment and stimulates regional economies via raised income. Beyond job creation, investment in human capital carries wider ramifications for regional advancement [12]. By furnishing local populations with specialized expertise pertinent to the fuel and energy sector, these investments bolster innovation, technological progress, and productivity in the region. This, in turn, leads to economic variety as new commercial sectors emerge or existing ones grow due to better access to energy resources. Nonetheless, it is vital to acknowledge that human capital investment in the fuel and energy complex may also lead to socio-economic inequalities within regions or nations [13, 14].

## 2. Positive externalities and key investment factors

Currently, the global community is confronted with the formidable task of tackling climate change and curtailing environmental harm, chiefly originating from the utilization of polluting energy resources. Consequently, it is apparent that fostering human capital development stands as a pivotal catalyst for beneficial transformation [18, 19]. Human capital, which includes attributes such as education, expertise, insights, and inventiveness, engenders a spectrum of beneficial externalities capable of transforming the energy industry and alleviating the detrimental environmental impact of fossil fuel use [20].

Table 1 that follows offers a summary of the positive externalities of human capital in the ongoing transition to the green energy that is centered along the three main pillars: i) human capital and innovation; ii) technological advancement; and iii sustainable practices (see Table 1).

**Table 1.** Positive externalities of human capital in the shift to the green energy

| Positive externalities | Description and explanation |
|---|---|
| | Human capital and innovation |
| Education and research | An educated populace engages in research and development activities, fostering the creation of clean energy solutions. |
| Skill diversity | Diverse skills within the workforce drive innovation and the development of clean energy technologies |
| | Technological advancement |
| Entrepreneurship and start-ups | Human capital fosters entrepreneurship, leading to the creation of start-ups specializing in clean energy technologies. |
| Knowledge transfer | Skilled individuals facilitate the transfer of knowledge, accelerating the adoption of clean energy technologies. |
| | Sustainable practices |
| Environmental awareness | An educated society is more environmentally conscious, leading to increased demand for clean energy sources. |
| Policy advocacy | Human capital can influence policymaking by advocating for renewable energy policies and environmental regulations. |

Source: Own results

Education and scholarly inquiry are pivotal in harnessing the beneficial externalities of human capital for the transition towards renewable energy sources. A well-educated society is more inclined to partake in research and development endeavors that focus on innovating in the realm of clean energy [21]. Academic and research institutions are at the forefront of conceiving new technologies and methods to enhance sustainable energy production [22]. A labor force that boasts a diverse array of competencies, including those in engineering, renewable energy technology, and environmental sciences, is better positioned to spearhead the growth and actualization of renewable energy initiatives [23].

Moreover, it is widely acknowledged that human capital stimulates entrepreneurial efforts and the emergence of start-ups dedicated to renewable energy technologies. These enterprises are instrumental in propelling technological advancements and competitive markets, which in turn yield more economical and efficient renewable energy options [24, 25]. In the academic and industrial sectors, knowledgeable individuals are crucial in expediting the dissemination of expertise and technology, thus hastening the embrace of renewable energy solutions [26].

Furthermore, it is evident that an educated populace tends to be more ecologically mindful, resulting in heightened demand for renewable energy and diminished reliance on polluting energy forms [27]. Consequently, human capital has the potential to shape policy by lobbying for the adoption of renewable energy legislation, carbon pricing, and environmental norms that favor the uptake of renewable energy.

Research consensus also indicates that among the essential elements affecting the socio-economic outcomes of investing in human capital are education and vocational training, job creation, knowledge dissemination, and the cultivation of social infrastructure [28, 29]. Table 2 lists the key factors influencing the socio-economic impact of investments in human capital (Table 2).

**Table 2.** Key factors of the socio-economic impact of investments in human capital

| Key factors | Description |
|---|---|
| Education and training | Quality education and training programs enhance workforce skills, promote innovation, and drive productivity, contributing to regional development |
| Job creation | Human capital investments lead to job creation, reducing poverty, improving living standards, and fostering economic growth through direct and indirect employment |
| Knowledge transfer | Investments facilitate knowledge sharing, transferring technological advancements and industry expertise to local communities, enhancing regional capabilities |
| Social infrastructure development | Companies contribute to social infrastructure like schools, hospitals, housing, and community centers, benefiting employees and positively impacting regional development |

Source: Own results

The educational attainment and skill level of personnel are key determinants of the socio-economic benefits derived from human capital investments. Firms within the fuel and energy sector are advised to prioritize comprehensive education and training for their workforce. Enhancing their technical skills will spur innovation and elevate productivity, ultimately contributing to regional economic advancement [30, 31].

The existing body of research corroborates that human capital investments catalyze job creation, which in turn exerts a notable socio-economic influence on localities. These investments not only create jobs within the energy companies but also spur employment indirectly via the supply chain and ancillary services. An upswing in employment is associated with reductions in poverty, enhancements in quality of life, and economic expansion [32, 33].

Furthermore, human capital investments are pivotal in the transfer of knowledge from corporations to the local populace. Disseminating technological innovations, exemplary practices, and specialized knowledge to community members aids in broadening their skill sets and cultivates entrepreneurial initiatives. This transference bolsters local capacities and paves the way for entrepreneurial ventures [34, 35].

Lastly, numerous studies indicate that corporations investing in human capital frequently contribute to the development of communal infrastructure, including educational institutions, healthcare facilities, residential projects, and community centers. These contributions serve not just the workforce but also have a wider positive impact on the socio-economic fabric of the regions where energy and fuel corporations operate [36-40].

## 3. Materials and methods

In the empirical part of our study, we employ bibliometric analysis that is based on a dataset of 1933 documents (represented by research papers, conference proceedings, and book chapters) indexed in the Web of Science (WoS) database between 1991 and 2023. We have selected the Web of Science (WoS) database as one of the most prestigious and complete abstract and citation databases that features a vast amount of research on this and related topics. From Figure 1 that follows, one can see the growing trend in the number of publications on the role and place of human capital in the energy sector (Figure 1).

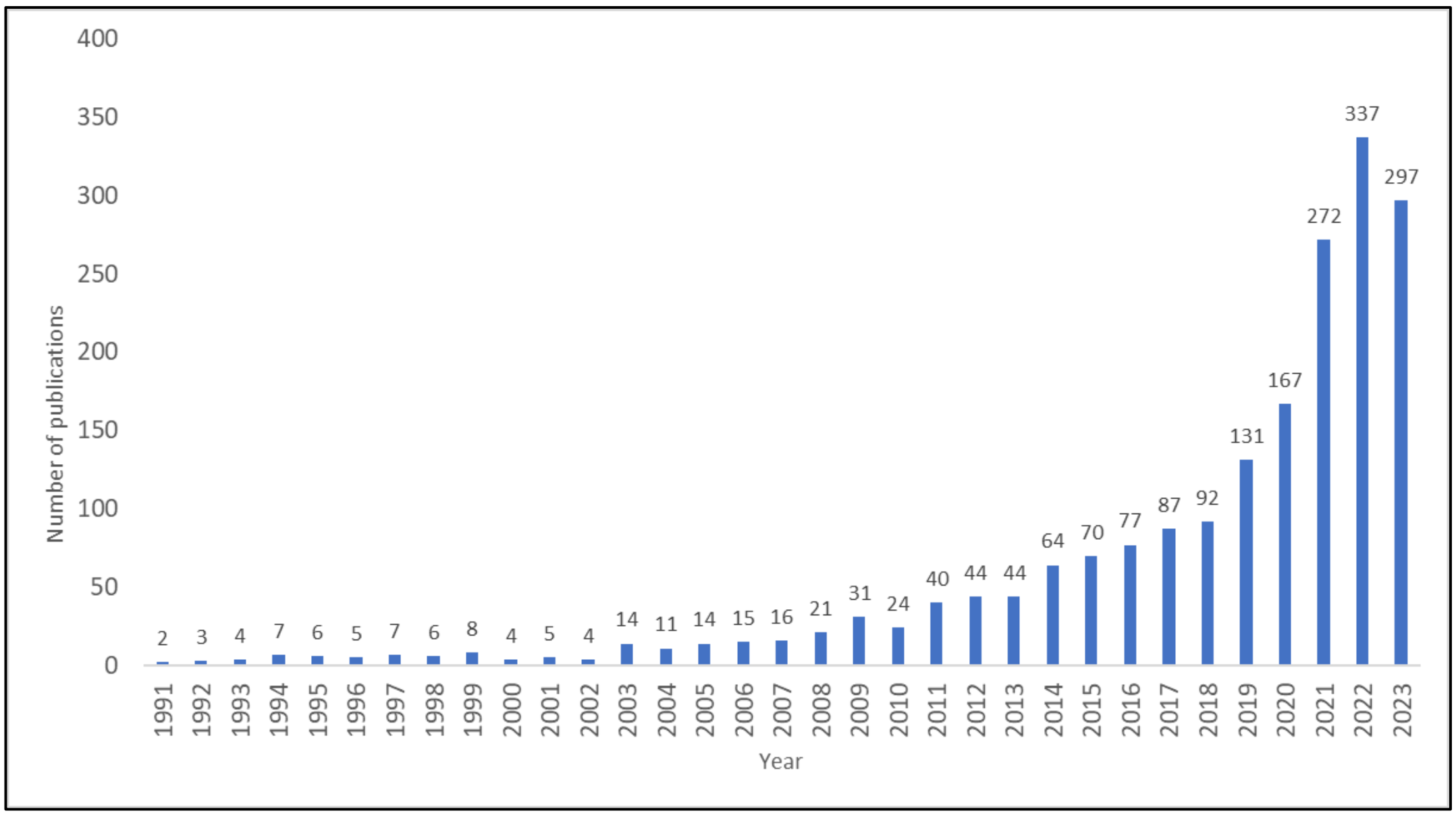

**Fig. 1.** Annual number of publications on human capital and energy sector (1991-2023).
Source: Own results based on WoS.

Using the set of data described above, we conducted a network cluster analysis of the textual and bibliometric inputs using VOSViewer software. Figure 2 that follows provides the description of our algorithm for the data selection, retrieval, processing, as well as the network analysis used in our study. The diagram delineates the methodological approach for data selection and network analysis in a stepwise fashion. Initially, the research phenomenon is defined, which informs the choice of the Web of Science (WoS) as the database for sourcing research publications. The search criteria are then established through two distinct queries: the first query utilizes the Google Trends toolkit to search for the terms "human capital" AND "energy sector" spanning the years 2004 to 2023, while the second query searches the WoS database for "human capital" AND "energy" covering a broader timeframe from 1991 to 2023.

The search within the chosen database is limited by specific criteria, including the scope (title, abstract, and keywords), the timeframe (1991 to 2023), language, the quality of the studies, and the type of document, which includes articles, conference proceedings, and book chapters. Following the establishment of search parameters, data extraction from the WoS database is conducted, leading to a general analysis of the WoS data. This analysis considers the distribution of documents by year, affiliations, journals, and institutions. The extracted data is then inputted into the VOSviewer software, a tool designed for constructing and visualizing bibliometric networks. The analysis of the results from VOSviewer includes network and overlay visualization, the identification of the number of clusters, types of clusters, and the total link strength. This step also involves discerning the implications of the network analysis. Finally, the process concludes with the synthesis of the findings to draw conclusions and implications from the study. This comprehensive approach ensures a systematic and rigorous analysis of the data, facilitating a robust understanding of the research phenomena (Figure 2).

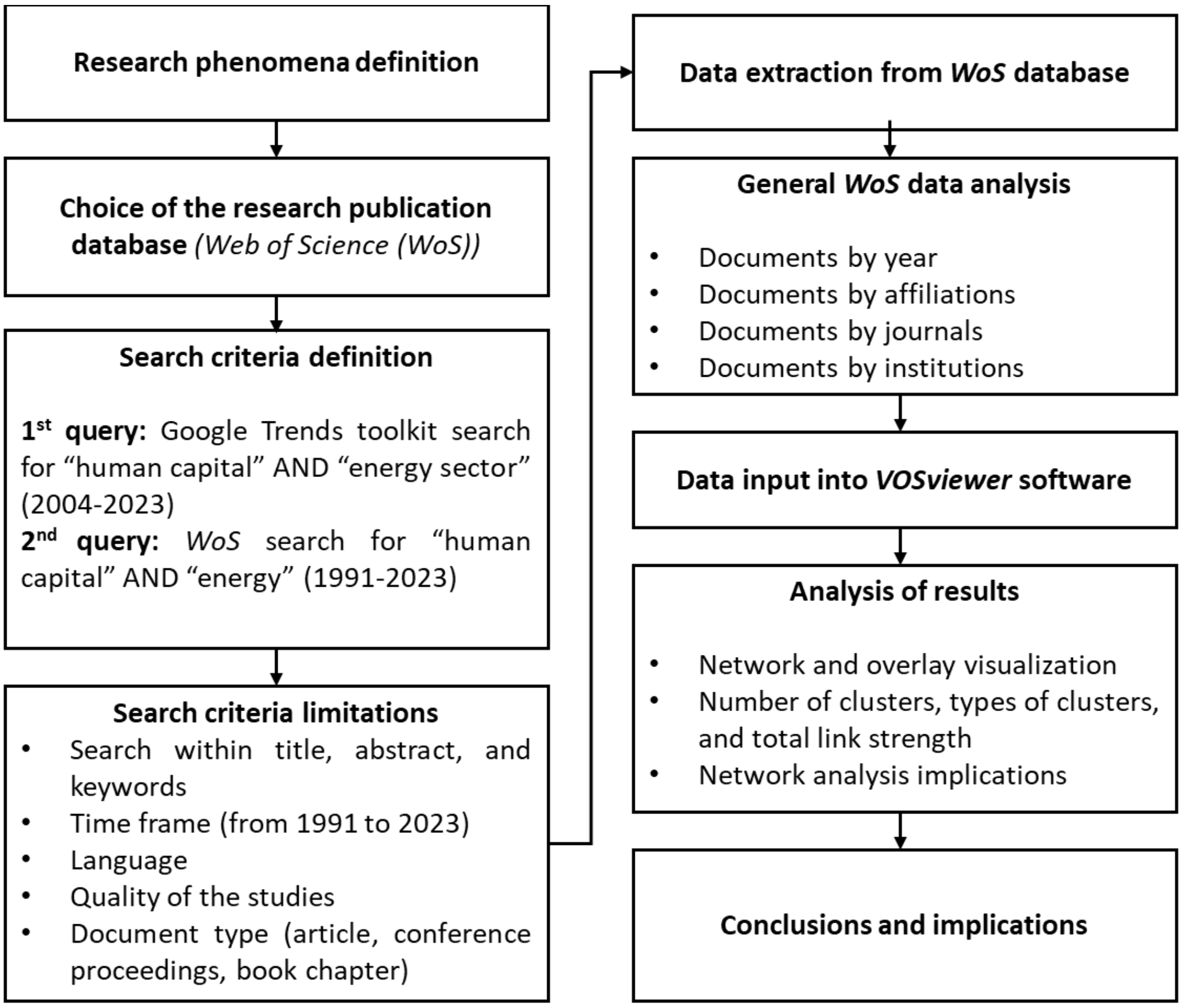

**Fig. 2.** Diagram of the data selection and network analysis algorithm.
Source: Own results

**4. Main results**

Figure 3 that is presented below presents the results from the online analysis using the Google Trends toolkit. Google Trends helps the researchers to identify the popularity of specific terms in various parts of the world measured by online searches.

It is apparent from Figure 3 that while the searches (measured by the Interest over Time (IoT), the metrics which assesses the search interest relative to the highest point on the chart for the given region and time (with 100 being the peak popularity and 0 showing that no data was available for the given region and time period) for "human capital" surpass those for "energy sector, there is still a clear correlation in both. In addition, while the search interest for "energy sector" appears to yield a steady and stable trend, the one for the human capital is growing (after a small decline between 2009 and 2018). All of them signal the existence of a certain interrelationship between these two concepts (see Figure 3 that follows).

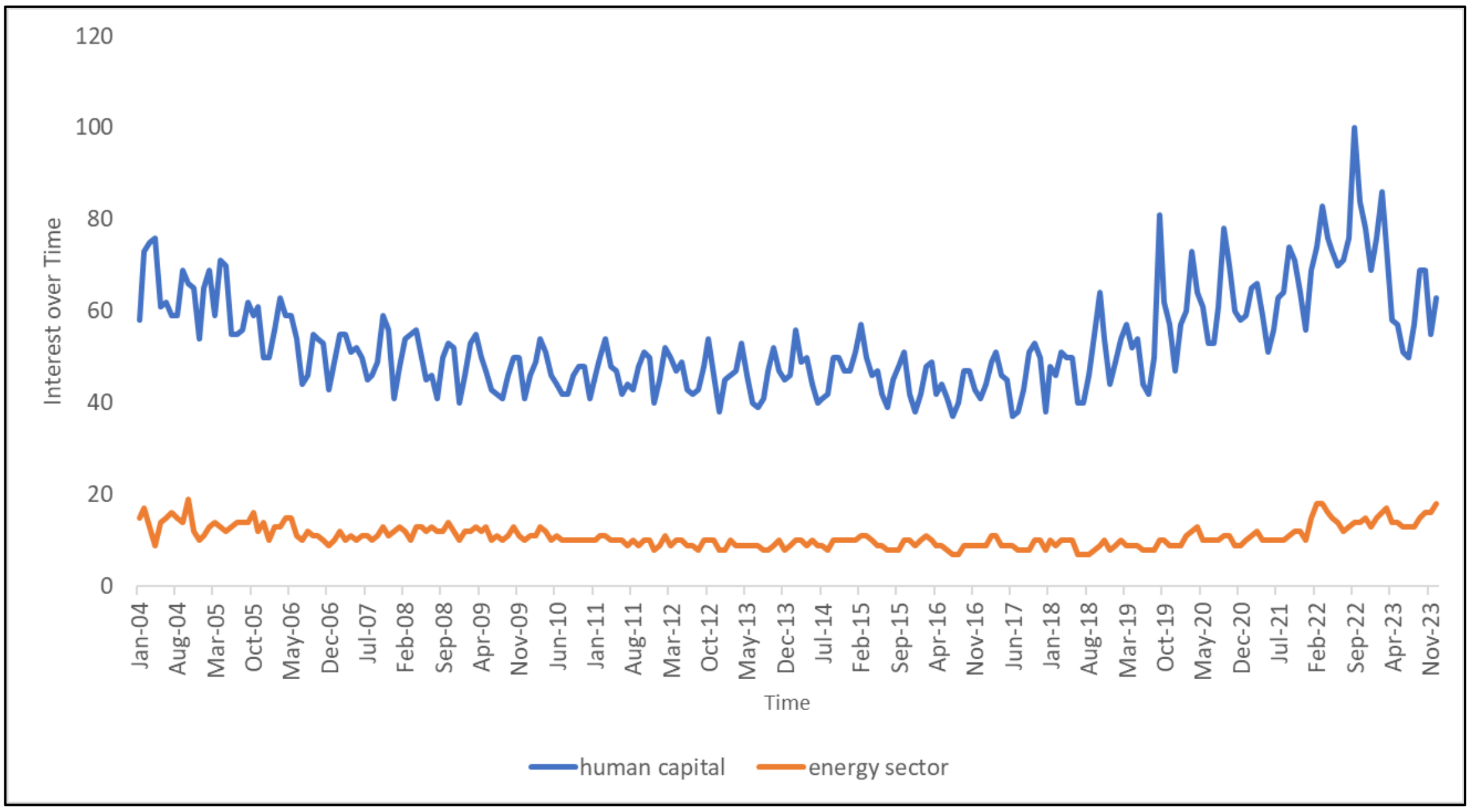

**Fig. 3.** Google Trends of the publications on "human capital" and "energy sector" (2004-2023). Source: Own results based on Google Trends

Table 3 that follows lists the top 10 cited articles covering human capital and energy published and indexed in WoS between 1991 and 2023.

**Table 3.** Top 10 most cited articles (1991-2023)

| Authors | Title | Number of citations |
|---|---|---|
| Ahmed et al. (2020) | Declarations for sustainability in higher education: becoming better leaders, through addressing the university system | 542 |
| Zafar et al. (2019) | The impact of natural resources, human capital, and foreign direct investment on the ecological footprint: The case of the United States | 407 |
| Hao et al. (2021) | Green growth and low carbon emission in G7 countries: How critical the network of environmental taxes, renewable energy and human capital is? | 353 |
| Pata and Caglar (2021) | Investigating the EKC hypothesis with renewable energy consumption, human capital, globalization and trade openness for China: Evidence from augmented ARDL approach with a structural break | 350 |
| Ahmed et al. (2020) | Linking urbanization, human capital, and the ecological footprint in G7 countries: An empirical analysis | 343 |
| Ren et al. (2021) | Digitalization and energy: How does internet development affect China's energy consumption? | 326 |
| Hassan et al. (2019) | Linking economic growth and ecological footprint through human capital and biocapacity | 323 |
| Khan et al. (2021) | How does fiscal decentralization affect CO2 emissions? The roles of institutions and human capital | 318 |
| Liang and Yang (2019) | Urbanization, economic growth and environmental pollution: Evidence from China | 302 |
| Jahanger et al. (2022) | The linkages between natural resources, human capital, globalization, economic growth, financial development, and ecological footprint: The moderating role of technological innovations | 267 |

Source: Own results based on Web of Science database [38-47]

Figure 4 that follows depicts the co-occurrence map based on the text data of the 1933 papers containing the keywords “human capital” and “energy” retrieved from WoS database. Our analysis has identified three main clusters (marked in red, green, and blue colors).

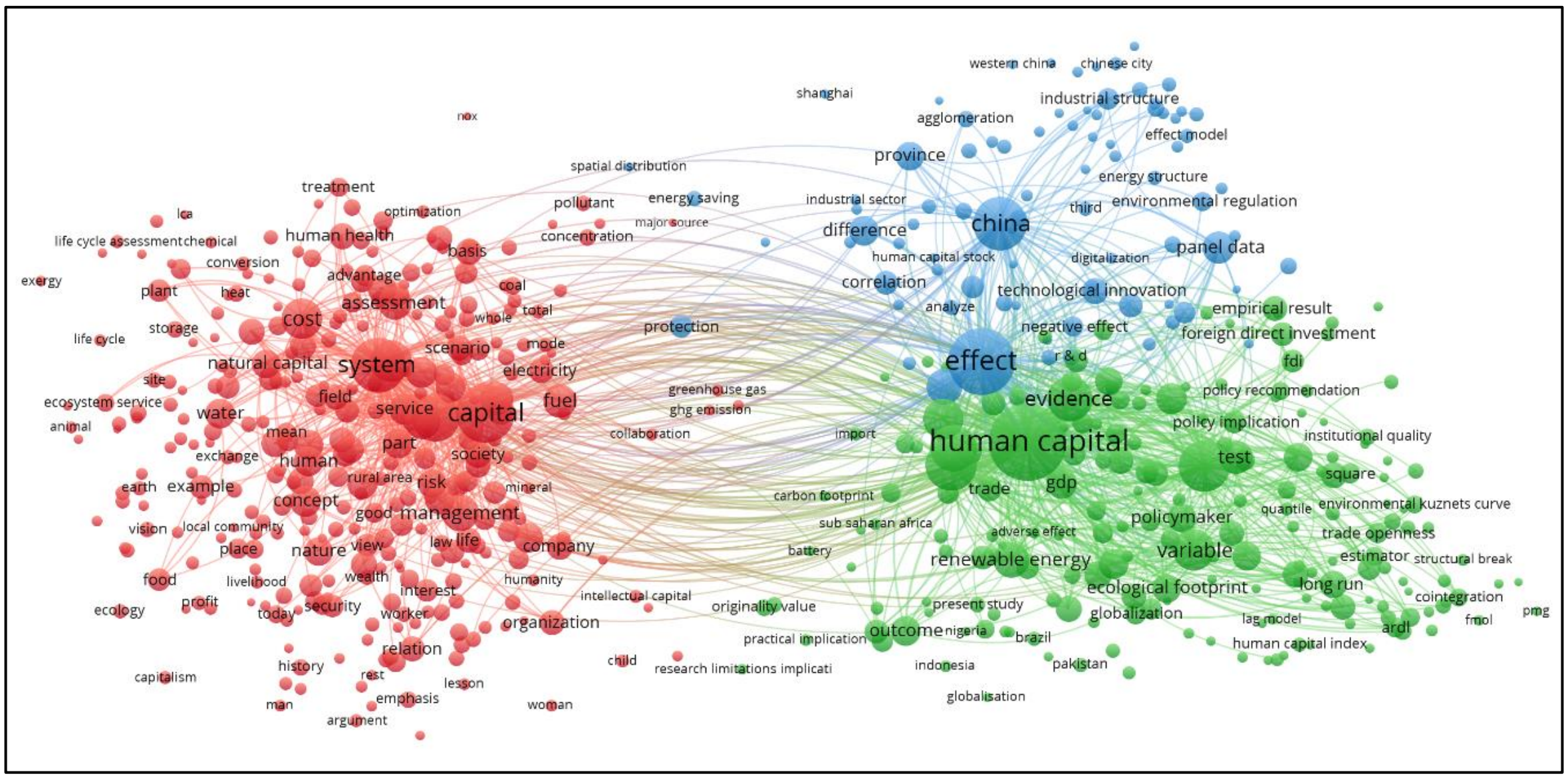

**Figure 4.** Co-occurrence map based on the text data of the 1933 papers containing the keywords “human capital” and “energy” retrieved from WoS database.
Own results based on VOSViewer v.1.6.15 software

In Figure 4, interconnections between clusters are apparent indicating interdisciplinary research efforts. The prominence of "human capital" in proximity to "outcome" and "renewable energy" could point to findings that investments in human capital are crucial for achieving positive outcomes in energy efficiency and the broader adoption of renewable energy technologies. The presence of terms like "policy implication" and "economic growth" close to "human capital" hints at the policy relevance of this research, emphasizing human capital's role in economic development within the energy sector.

Overall, the map indicates a multidisciplinary dialogue within the literature, where economic, environmental, and policy dimensions converge around the pivotal concept of human capital in the context of energy research.

In addition, Figure 5 that follows below depicts the bibliographic map based on the bibliometric data (co-authorship, keyword co-occurrence, citation, bibliographic coupling, or co-citation map) of the 1933 papers containing the keywords “human capital” and “energy” in WoS database (see Figure 5).

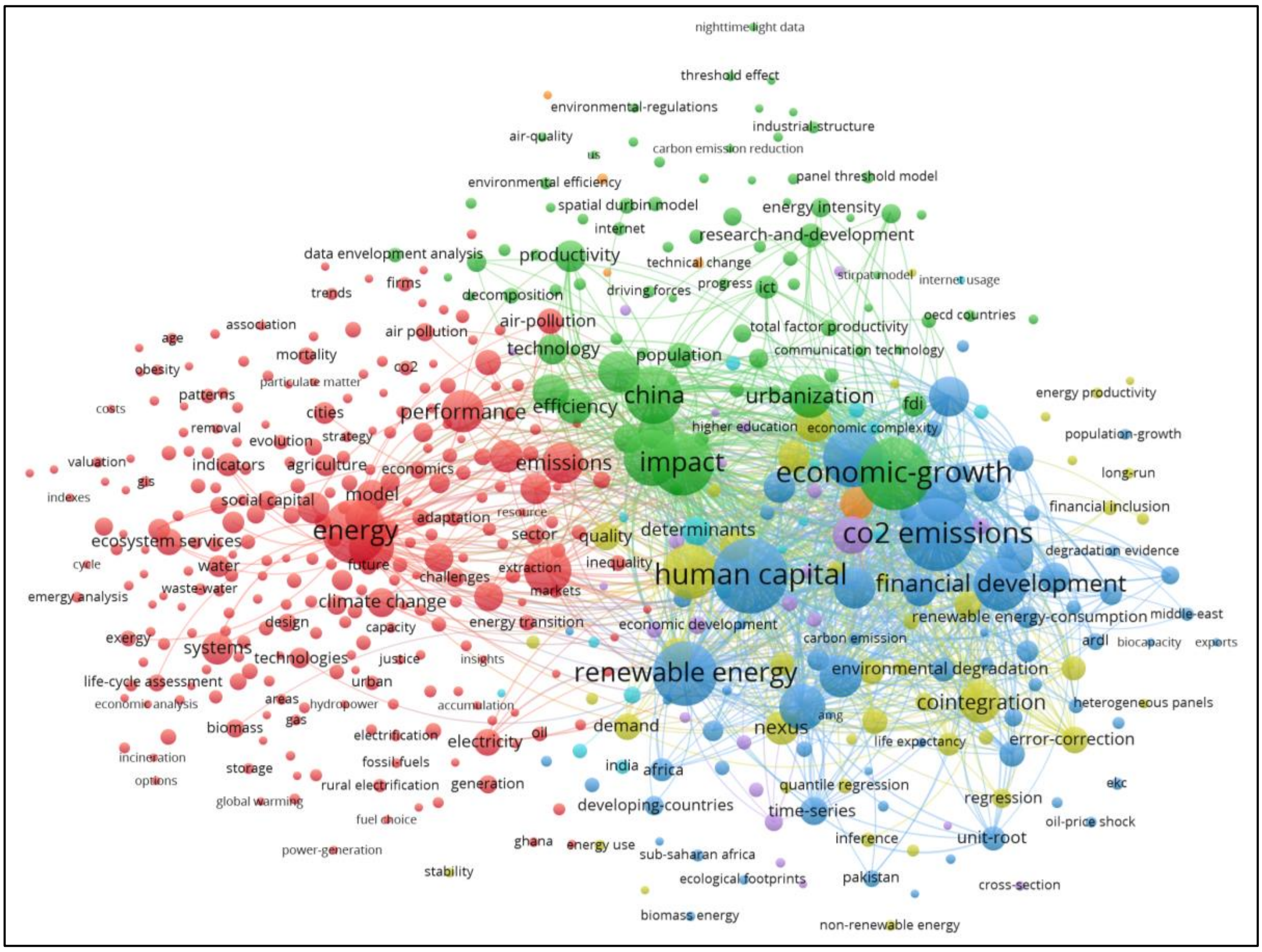

**Figure 5.** Bibliographic map based on the bibliometric data (co-authorship, keyword co-occurrence, citation, bibliographic coupling, or co-citation map) of the 1933 papers containing the keywords “human capital” and “energy” in WoS database.
Own results based on VOSViewer v.1.6.15 software

The map displays a dense network of terms clustered into groups, which are color-coded, indicating different thematic focuses within the combined realms of human capital and energy.

The red cluster seems to concentrate on foundational and systemic aspects, with terms like "model," "system," "change," and "energy" suggesting a focus on conceptual and systemic analyses of energy sectors. There's an apparent emphasis on systemic changes and models related to energy, which may involve discussions on lifecycle, efficiency, and sustainability.

The blue cluster features terms like "economic growth," "urbanization," "CO2 emissions," and "China," pointing to a research trend that examines the economic dimensions of energy, particularly in relation to urban development and environmental impact. This cluster indicates a significant interest in the intersection of energy consumption and economic development, especially in rapidly developing and urbanizing contexts.

The green cluster centers around "human capital" and "renewable energy," surrounded by terms such as "financial development," "environmental degradation," and "renewable energy consumption." This suggests that the research is examining the influence of human capital on the adoption and development of renewable energy technologies and its implications for financial markets and environmental health.

The interconnectedness of clusters indicates that the research field is highly interdisciplinary, integrating human capital concerns with energy policy, environmental impact, and economic factors. The prominence of "human capital" in the context of "renewable energy" and "CO2 emissions" suggests an analysis of how education and skill development can contribute to environmental sustainability and energy efficiency.

All in all, the network diagram map suggests a vibrant and interconnected field of study where the expertise and attributes of the workforce and human capital are considered crucial to advancing renewable energy technologies, improving environmental outcomes, and driving economic growth in the context of energy use and policy.

## 5. Conclusions and implications

Overall, our results stemming from both the literature review and the bibliometric network analysis demonstrate that human capital investment significantly influences sustainable socio-economic growth within the energy sector. Many researchers unanimously agree that by allocating resources to the education and training of their workforce, energy firms can enhance the competencies and expertise of their employees. Such enhancements lead to heightened productivity, greater operational efficiency, and regional economic expansion. Moreover, these investments cultivate a labor force adept at navigating shifts in technology and market dynamics. Additionally, lots of studies are in accord about the fact that human capital investment represents a catalyst for regional innovation and technological progress, as it empowers employees to engage in research and development, thus fostering novel approaches and solutions that have broad regional benefits.

In addition, human capital investment also carries substantial social benefits. It results in improved living conditions through job creation, increased income, and access to vital services, which collectively work to diminish economic disparities and elevate the region's standard of living. In essence, leveraging human capital is key to promoting sustainable development in locales where energy and fuel industries are prominent.

Within this context, it appears important to emphasize that human capital's benefits transcend individual and economic advancement. Within the environmental sustainability framework, it is pivotal for spurring innovation, technological progression, and the embrace of renewable energy practices. As the global community confronts climate change and aims for a shift towards more environmentally friendly energy systems, the strategic enhancement of human capital becomes essential. It is incumbent upon policy architects, key stakeholders, educational entities, and industry frontrunners to recognize human capital's instrumental role in transitioning from traditional to green energy consumption, which is vital for a healthier and more sustainable global environment. A concerted effort is required to deepen and broaden support for skill and knowledge development, alongside investments in human capital from both private entities and public institutions.

**Funding:** The study was supported by the Russian Science Foundation grant No. 23-28-01768, https://rscf.ru/project/23-28-01768.

**Conflicts of Interest:** The authors declare no conflict of interest.